\documentstyle[11pt,newpasp,twoside]{article}
\reversemarginpar
\begin{document}
\title{Retardation of Gravity in Binary Pulsars}
\author{Sergei M. Kopeikin}
\affil{Department of Physics and Astronomy, University of Missouri-Columbia, Columbia, Missouri, 65211, USA}
\baselineskip=11.5pt
\begin{abstract}
We study the effect of retardation of gravity in binary pulsars. It appears in pulsar timing formula as a periodic excess time delay to the Shapiro effect. The retardation of 
gravity effect can be large enough for observation in binary pulsars with the nearly edgewise orbits and relatively large ratio of the projected semimajor axis, $x$, to 
the orbital period of the pulsar, $P_b$. If one succeeds in measuring the retardation of gravity it will give further experimental evidence in favor of General Relativity. 
\end{abstract}
\section{Introduction}
Timing of binary pulsars is one of the most important methods of testing
General Relativity both in the weak and strong gravitational field regimes (Manchester \& Taylor 1977; Damour \& Taylor 1992; Taylor 1994; Kaspi, Taylor,\& 
Ryba 1994; Stairs et al. 1998; Van Straten et al. 2001). Such an opportunity exists because of the possibility to 
measure in some of the binary pulsars a sufficiently large number of classical and relativistic  
parameters of the pulsar's orbital motion. Each of the parameters
depends on masses of orbiting stars and orbital characteristics in a different functional way. Consequently, if
three or more of these parameters can be measured in addition to five Keplerian parameters (Manchester \& Taylor 1977), the overdetermined
system of the parameter equations can be used to test the gravitational theory (Damour \& Taylor 1992; Kopeikin \& Potapov 1994).

Especially important for this test are binary pulsars on the
orbits visible nearly edge-on. In such systems
masses of orbiting stars can be easily determined by measuring the range, $r$, and shape, $s$, of the Shapiro time delay (Shapiro 1967; Blandford \& Teukolsky 1976) 
in the propagation of the radio pulses from the pulsar to the observer. 
Perhaps, the most famous examples of the nearly edge-on binary 
pulsars are PSR B1855+09 and PSR B1534+12. The sine of the inclination angle $i$ 
of the orbit of PSR B1855+09 to the line of sight makes up a value of about 
0.9992 and the range parameter of the Shapiro effect reaches 1.27 $\mu$s 
(Kaspi et al. 1994). The corresponding quantities for PSR B1534+12 are
$\sin i=0.982$ and $r=6.7 \mu$s respectively (Stairs et al. 1998). 

All binary pulsars emit
gravitational waves, a fact which was confirmed with the precision of 
about 
$0.3\%$ by Joe
Taylor and collaborators (Taylor \$ Weisberg 1989; Taylor 1994). New achievements in technological
development of data acquisition systems and continuous upgrading the largest radio telescopes (Bailes 2002) extend our
potential to measure with higher precision the time-of-arrival of pulsar's radio signals. Therefore, new classic and relativistic effects can be measured in a very near 
future. 

Among them the influence of the velocity-dependent terms in the metric tensor of a binary pulsar on propagation of radio signals from the pulsar to the observer is of 
a special interest. These terms produce an 
additional 
effect in timing formula that reveals itself as a small periodic excess to the 
range and shape of the Shapiro time delay making it dependence on the orbital phase of the binary system be more complicated function of time. The mathematical 
properties of the effect under discussion were thoroughly discussed by Kopeikin \& Sch\"afer (1999). Later, Kopeikin (2001) associated this effect with the retarded 
nature of the Einstein field equations. Thus, the excess to the range and shape of the Shapiro effect can be interpreted as a direct consequence of the fact that gravity 
propagates with the finite speed that can be actually measured in the nearly edge-on relativistic binary pulsars. This paper is designed primarily for experimenters 
to give them further insight to the problem of detection of the retardation of gravity in binary pulsars.
\section{Gravitational Time Delay in Binary Pulsars}

In this section we present the gravitational time delay formula which includes, besides of the well known Shapiro's logarithm, all linear corrections for the 
velocities of the pulsar and its companion. 
The original idea of the derivation of the gravitational time delay in the
{\it static} and 
{\it spherically symmetric} field of a self-gravitating body belongs to
Irwin 
Shapiro (1967). Regarding binary pulsars the  
Shapiro time delay has been computed 
by Blandford \& Teukolsky (1976) under the assumption that gravitational field is weak and static everywhere. This was actually an unjustified approximation 
because the pulsar and its companion move around the barycenter of the binary as radio signal from the pulsar propagates towards observer. We have developed 
(Kopeikin \ Sch\"afer 1999; Kopeikin \& Mashhoon 2002) a novel theory of propagation of light in time-dependent gravitational fields that allows to treat the 
relativistic 
time delay rigorously and account for all effects caused by the {\it 
non-stationary} 
part of the gravitational field of a binary pulsar, that is to
find all relativistic 
corrections of order $v/c$, $v^2/c^2$, etc. to the Blandford-Teukolsky formula, where $v$ denotes a characteristic velocity of
bodies in the binary pulsar and $c$ should be understood as the speed of gravity (Kopeikin 2001). Of course, in general relativity the speed of gravity is numerically 
the same as the speed of light. However, it is the speed of gravity which enters the equation describing the propagation of gravity from moving pulsar and its 
companion to a radio signal emitted by the pulsar. 

The original formula for the gravitational time delay is quite complicated (see Eq. (51) in (Kopeikin \& Sch\"afer 1999) for more detail). It depends on the position of 
pulsar and its companion at the retarded time taken by gravity to propagate from these bodies to observer (Kopeikin 2001). One can approximate the original formula 
by making use of the expansion with respect to the small parameter $v/c$ so that it assumes the following form (Kopeikin \& Sch\"afer 1999)
\begin{equation}
\label{ert}
\Delta_g=-{2Gm_c\over c^3}\left(1-{1\over c}\,{\bf k}\cdot{\bf v}_c\right)
\ln\left[r+{\bf k}\cdot {\bf r}-{1\over c}\,\left({\bf k}\times{\bf v}_c \right)\cdot
\left({\bf k}\times{\bf r}\right)\right],
\end{equation}
where ${\bf k}$ is the unit vector from the pulsar to observer, ${\bf r}$ is the radius-vector from pulsar's companion to the pulsar, $r=|{\bf r}|$, ${\bf v}_c$ is the 
velocity of the companion with respect to the barycenter of the binary system, $"\cdot"$ and $"\times"$ denote the Euclidean scalar and vector products.
\section{Parametrization of the Gravitational Time Delay}

One notices the presence of the relativistic terms of order $1/c$ in front of and in the argument of the logarithmic function describing the Shapiro time delay in binary 
pulsar. Both these corrections are directly related to the effect of propagation of gravity in General Relativity with the finite speed equal to the speed of light. In 
other alternative theories of gravity these relativistic terms would not have such a simple form. Thus, their measurement give an additional test of general relativity 
being independent of all previous tests. Measurement of the relativistic terms under discussion is, however, a great challenge for pulsar timing community. The 
matter is that the additional effects are proportional to the ratio of $v/c$ that amounts to magnitude $10^{-3}$ in some relativistic binaries. The ratio $v/c$ can be 
larger than $10^{-3}$ only in the case of ultra-relativistic binary systems with massive stars and very short orbital periods. Finding such relativistic systems should 
be conceived as a first-priority task of pulsar searches (Jouteux et al. 2002).

The term in front of the logarithm modulates the range parameter of the Shapiro time delay (see Eq. (5) below). The amplitude of this modulation can reach for the 
known binary pulsars 10 nanoseconds which is probably too small to detect presently. The relativistic term in the argument of the logarithmic function is more 
interesting. This term can be important for relativistic binary pulsars with nearly edge-on orbits where it changes the shape parameter of the Shapiro time delay 
making it dependent on the orbital phase. One can show that Eq. (1) can be reduced to the form looking similar to the Blandford-Teukolsky parametrization of the 
Shapiro effect. Specifically, after straightforward calculations of the scalar and vector products in Eq. (1) by making use of the Keplerian orbital parametrization of 
the two-body problem (Klioner \& Kopeikin 1994), one has 
\begin{equation}\hspace{-0.04cm}
\label{par}
\Delta_g=-2r\ln\left\{1-e\cos(u+\varepsilon)  - s\left[
\sin\omega(\cos u-e)+\sqrt{1-e^2}\cos\omega\sin u\right]\right\},
\end{equation}
where $e$ is the orbital eccentricity of the binary, $\omega$ is the argument of the periastron, and $u$ is the eccentric anomaly relating to the time of emission, $T$, 
and the instant of the first passage of the pulsar through the periastron, $T_0$, by the Kepler transcendental equation
\begin{equation}
\label{k}
u-e\sin u={2\pi\over P_b}\left(T-T_0\right)\,.
\end{equation}
The other parameters entering Eq. (2) are  
\begin{eqnarray}
\varepsilon&=&\frac{2\pi}{\sin i}
\frac{x}{P_b}\frac{m_p}{m_c}\,,\\
r&=&\frac{G m_c}{c^3}\left[1 -\frac{\varepsilon\sin i}{\sqrt{1-e^2}}\,F(u)\right]\,,\\
s&=&\sin i\left[1 +\frac{\varepsilon\sin i}{\sqrt{1-e^2}}\,
F(u)\right]\,,
\end{eqnarray}
where $m_p$ and $m_c$ are masses of the pulsar and its companion respectively, $x=a\sin i/c$ is the orbital semimajor axis of pulsar's orbit projected on the line of 
sight and measured in seconds, and function
\begin{equation}
F(u)=e \cos\omega+\frac{(\cos u-e)\cos\omega-\sqrt{1-e^2}
\sin\omega\sin u}{1-e\cos u}\,.
\end{equation}
In the case of a nearly circular orbit, when $e\simeq 0$, Eq. (2) is simplified
\begin{equation}
\Delta_g=-\frac{2G m_c}{c^3}\left(1-\varepsilon\sin i\cos\phi\right)\ln\left(1-\sin i\sin\phi-{1\over 2}\,\varepsilon\sin^2 i\sin 2\phi\right)  ,
\end{equation}
where $\phi=u+\omega$ is the orbital phase. One notices that the argument of the logarithmic function is modulated by the term having a double orbital frequency.

In Eqs. (4)--(6) parameter $\varepsilon$ is the new (constant) relativistic parameter and
$r$ and $s$ are the range and shape parameters of the Shapiro time delay. It is worth noting that both of these parameters are no longer constant but depend of the 
orbital phase. This distorts the amplitude and shape of the logarithmic curve. Measurement of this distortion allows to make a judgment about how fast gravity 
propagates with respect to electromagnetic waves. We also would like to emphasize that the magnitude of the new effects described in this paper can be large enough only 
in the binary systems where mass and orbital velocity of pulsar's companion are reasonably big. The binary can be visible edge-on but if pulsar's companion is a 
low-mass star the detection of the retardation of gravity effect is questionable. 

Similar study applied to the case of light propagating in the solar system reveals that the retardation of gravity effect can be also detected in the event of a close 
passage of massive Jupiter near a bright quasar (Kopeikin 2001). 

\end{document}